\documentclass{iaus}
\usepackage{graphicx}
\usepackage{natbib}

\def\apj{ApJ}%
\def\mnras{MNRAS}%
\title[Ecology of galaxy stellar populations] 
{Ecology of galaxy stellar populations from optical spectroscopic surveys}

\author[Anna Gallazzi]   
{Anna Gallazzi$^1$}

\affiliation{$^1$Max-Planck-Institut f\"ur Astronomie, K\"onigstuhl 17, 69117, Heidelberg,
Germany \\ email: {\tt gallazzi@mpia.de}}

\pubyear{2010}
\volume{262}  
\pagerange{119--126}
\jname{Stellar Populations -- Planning for the Next Decade}
\editors{G. Bruzual \& S. Charlot, eds.}
\begin{document}

\maketitle

\begin{abstract}
The age and chemical composition of the stars in present-day galaxies carry important clues about their
star formation processes. The latest generation of population synthesis models have allowed to derive age
and stellar metallicity estimates for large samples of low-redshift galaxies. After reviewing the main
results about the distribution in ages and metallicities as a function of galaxy mass, I will concentrate
on recent analysis that aims at disentangling the dependences of stellar populations properties on
environment and on galaxy stellar mass. Finally, new models that predict the response of the full spectrum
to variations in [$\alpha$/Fe] will allow us to derive accurate estimates of element abundance ratios
and gain deeper insight into the timescales of star formation cessation.  
\end{abstract}

\firstsection 
\section{Introduction}

The age and chemical composition of the stellar populations in galaxies, together with galaxy mass, are
key ingredients to uncover galaxy formation and evolutionary paths. Estimates of stellar populations
parameters are derived by interpreting detailed spectral information, such as absorption features, on the
basis of stellar population synthesis (SPS) models. Our ability of interpreting galaxy spectra has greatly
improved with the development of SPS models that i) predict the full spectrum of simple stellar
populations (SSPs) at medium/high resolution, allowing to adjust the models to the data quality rather than the
other way round, and ii) have a better coverage of the stellar parameters space, allowing the
interpretation of a broader range of stellar populations
\citep{1999ApJ...513..224V,2003MNRAS.344.1000B,2007MNRAS.382..498C}.

Together with the development of SPS models and spectral fitting techniques, the statistical power of
large spectroscopic surveys has allowed to put on a firm ground our understanding of the stellar
populations in nearby galaxies. In this contribution I will briefly review results on the dependence of
stellar populations properties on galaxy mass, as obtained from the analysis of Sloan Digital Sky Survey
(SDSS) galaxy spectra. I will then discuss to what extent these relations are shaped by the environment in
which galaxies reside, as recently analysed in Pasquali et al (2009, submitted) combining the SDSS DR4
catalogue of stellar ages and metallicities with the \cite{2007ApJ...671..153Y} group
catalogue.  I will conclude with a brief outlook on new population synthesis models that allow accurate
estimates of $\alpha$-element abundance ratios.
  
\section{The ages and metallicities of present-day stellar populations}
\subsection{Estimates of stellar populations parameters}\label{sec1}
Estimates of stellar metallicity, luminosity-weighted age and stellar mass have been derived for the
spectroscopic sample of SDSS DR4 by interpreting the strength of a set of absorption features including
the 4000\AA~break, the Balmer lines and [$\alpha$/Fe]-insensitive Mg and Fe composite indices. A library
of galaxy spectral features is generated by convolving \cite{2003MNRAS.344.1000B} SSPs with a
comprehensive set of Monte Carlo star formation histories (SFHs). Following a Bayesian statistical
approach, the likelihood distribution of each parameter has been computed by comparing the observed
absorption features with those predicted by the full library.

The uncertainties on physical parameters estimates depend on the spectral signal-to-noise (S/N) and on the
galaxy spectral type or SFH. A careful analysis of the statistical uncertainties on stellar M/L, arguably
the easiest parameter to constrain, indicates two interesting points for model development and spectral
analysis: i) for galaxies with old stellar populations and spectra of good S/N statistical uncertainties
of 0.03~dex could in principle be reached, indicating the need for continuous model development and
control of systematics; ii) on the other hand for galaxies with recent bursts of star formation the
uncertainties are limited by the large intrinsic scatter in physical parameters at fixed index strength:
it is thus crucial to properly model the complexity of SFHs in the Universe \citep{GB09}.
\subsection{Dependence on stellar mass}\label{sec2}
In our previous studies, based on a sample of SDSS DR2 galaxies at $z<0.22$, we have analysed the
distribution of galaxies in stellar metallicity--age--stellar mass. In the mean, low-mass galaxies are
young and metal-poor while high-mass galaxies are old and metal-rich. There is a rapid transition between
these two regimes at stellar masses around $10^{10.5}M_\odot$ \citep{2005MNRAS.362...41G}. This
corresponds to the transition regime in other galaxy properties
\citep[e.g.,][]{2003MNRAS.341...54K,2004ApJ...600..681B}, and it is where the scatter in stellar
populations parameters is largest. 

Early-type galaxies, which contain the majority of baryons and metals in stars today \citep{2008MNRAS.383.1439G}, dominate above the transition mass where the relations between
metallicity/age and mass flattens. They follow a tight color-magnitude relation (CMR) which is driven by
an increase with stellar mass of both stellar age, total metallicity and element abundance ratios
\citep{2006MNRAS.370.1106G}, suggesting that the stars in massive early-type
galaxies have formed earlier and on shorter timescales than less massive ones. 
\subsection{Dependence on environment: centrals versus satellites}\label{sec3}
The bottom line of the above results is that galaxy SFH is primarily determined by galaxy mass. The
scatter in physical parameters at fixed mass, however, indicates underlying second-order dependences. The
star formation activity in galaxies is also sensitive to the environment in which galaxies reside, as
witnessed by the SFR-density relation \citep[e.g.,][]{2002MNRAS.334..673L} and by the differences in
scatter and zero-point of the  CMR of early-type galaxies in different environments
\citep[e.g.,][]{2005ApJ...621..673T,2006MNRAS.370.1106G}. These environmental dependences are however
subtle, once the primary dependence on stellar mass is removed. Moreover the signal may depend on the
environmental estimator adopted, such as e.g. galaxy number densities in apertures which have the
disadvantage of being sensitive to environment itself and not directly comparable to the more physical
description in terms of host dark matter halos of galaxy formation models.

\begin{figure}
\begin{center}
 \includegraphics[width=6.5truecm]{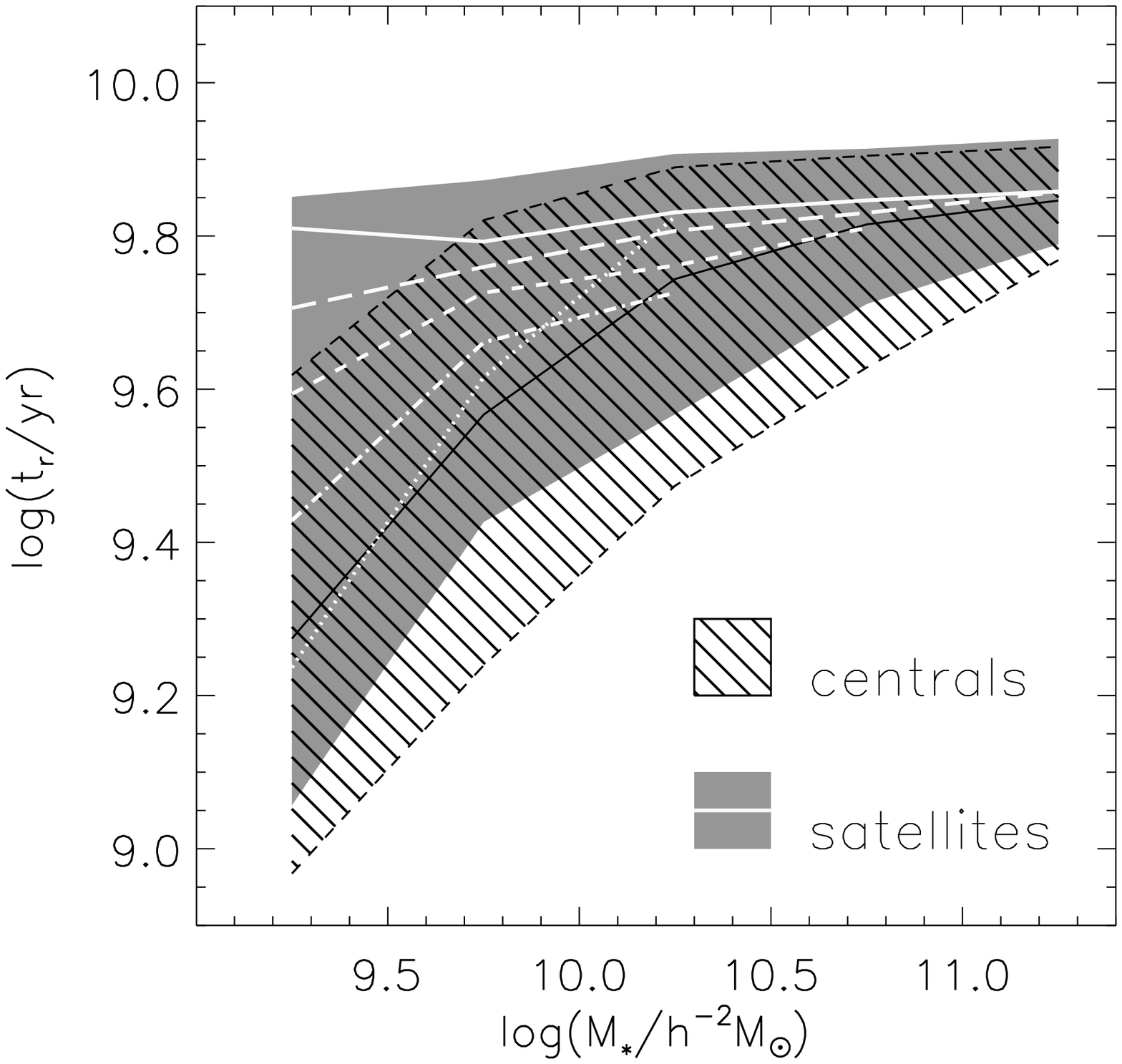} 
 \includegraphics[width=6.5truecm]{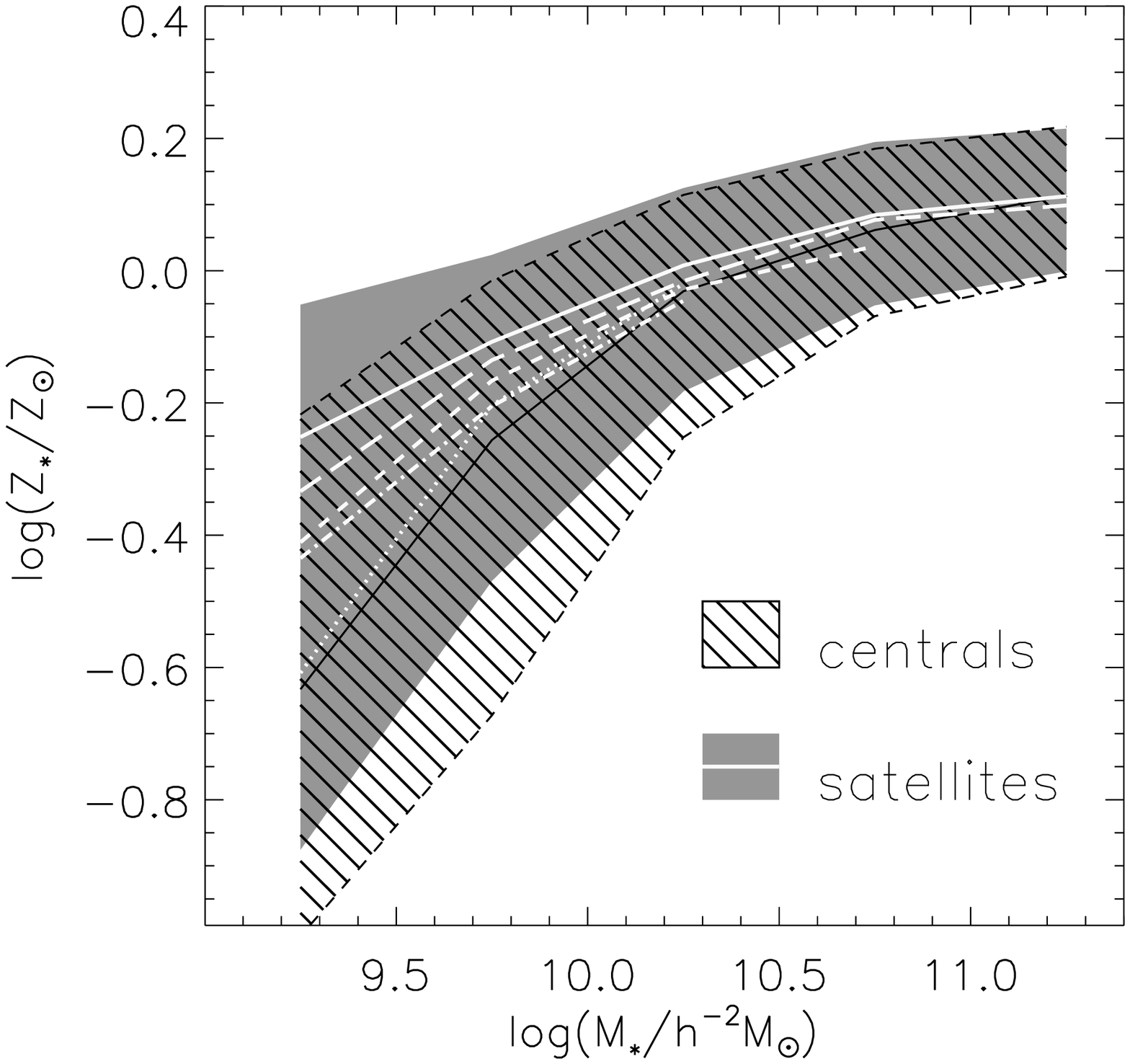}
 \caption{Luminosity-weighted age ({\it Left}) and stellar metallicity ({\it Right}) as a function of
 stellar mass. The full 16\%-84\% range of the distributions is shown. Galaxies are separated into
 centrals (hatched region) and satellites (grey shaded region). The median relations for satellites
 galaxies hosted by halos of different mass are shown by the white lines ($\rm M_{halo}/h^{-1}M_\odot$: 14--15
 (solid), 13--14 (long dashed), 12.5--13 (dashed), 12--12.5 (dot-dashed), 11--12 (dotted)). Figure adapted
 from Pasquali et al. (2009, submitted).}
   \label{fig1}
\end{center}
\end{figure}

In Pasquali et al. (2009) we take advantage of the group catalogue by \cite{2007ApJ...671..153Y} which assign a dark matter mass to galaxy groups and classify their members
into centrals and satellites. We complement the galaxy group catalogue with the stellar metallicity and
age estimates computed for SDSS DR4. We then analyse the dependence of stellar age and metallicity on
stellar mass, distinguishing centrals and satellites and as a function of host halo mass.

The left panel of Fig.~\ref{fig1} shows the relations between luminosity-weighted age and stellar mass
for central (hatched region) and satellite galaxies (grey shaded region). The relations between stellar
metallicity and mass are shown in the right-hand panel. At stellar masses $M_\ast>10^{10.5}M_\odot$
centrals and satellites have the same mean ages and metallicities. At lower stellar masses instead,
satellite galaxies are older and metal-richer than equally-massive centrals. Moreover, we note in the
satellite population an excess of old, metal-rich galaxies at low M$_\ast$ and a lack of young metal-poor
galaxies at nearly all masses. 
We reproduce the result found in other studies that at all scales, centrals are older and metal-richer
than satellites that live in equally-massive halos. This likely reflects the general trends of
age/metallicity with mass and the fact that in a given halo centrals are more massive than satellites. We
also find that, while the stellar populations of central galaxies are nearly independent of halo mass,
this is not the case for satellite galaxies. The white lines of different style in Fig.~\ref{fig1} show
the age/metallicity--mass relations for satellite galaxies in halos of different mass. Below
$M_\ast\sim10^{10.5}M_\odot$, at fixed stellar mass the stellar populations of satellite galaxies depend
on halo mass: satellites in more massive halos are older and metal-richer than those in less massive
halos.

At high masses, the SFH of galaxies is likely regulated by internal feedback processes regardless of
environment. At low masses instead the SFH of satellite galaxies is affected by environmental
processes, whereby their stellar populations differ from those of equally-massive centrals by an
amount that increases with halo mass.

\section{Element abundance ratios}\label{sec4}
The results discussed above indicate that satellite galaxies are affected by the environment onto which
they are accreted through mechanisms that quench star formation and prevent further
replenishment of gas via inflows. It remains to be determined on which timescale the SF in satellite
galaxies is quenched and any possible dependence on halo mass. Important clues about SF timescales can be
gained from the abundance of $\alpha$-elements with respect to Fe ([$\alpha$/Fe]). Previous results
suggest that the [$\alpha$/Fe] of isolated early-type galaxies is the same as for cluster galaxies
\citep{2005ApJ...621..673T}. A careful definition of environment (such as the one
introduced in the previous section) and accurate measures of [$\alpha$/Fe] are necessary to confirm or
disprove these findings.

Theoretical SPS models that predict the full spectrum of stellar populations with different [$\alpha$/Fe]
are now available \citep{2007MNRAS.382..498C}. Recently, \cite{2009MNRAS.398L..44W} have developed `differential' stellar populations models that exploit the
predictive power of fully theoretical models, calibrated onto semi-empirical solar metallicity SSPs. In
this way it is possible to predict the pixel-by-pixel absolute flux (as opposed to absorption
indices only) as a function of both [Fe/H] and [$\alpha$/Fe].

We are now in a position to predict spectra for stellar populations with variable [$\alpha$/Fe] (in
addition to metallicity) and with complex SFHs, and exploit the high resolution of the models to: i)
convolve them to match the effective resolution of the data (including velocity dispersion broadening),
ii) use the redundancy of the full spectral information for low-quality data, in addition to individual
absorption indices.

We have generated a Monte Carlo library of SFHs based on the new differential models. Adopting the same
Bayesian approach used before (see Section~\ref{sec1}) we can derive  consistent estimates of
$\alpha$-element abundance ratio together with luminosity-weighted age,  metallicity and mass-to-light
ratios, and their associated uncertainties. The analysis of a well-defined sample of early-type
galaxies will enable us to validate the model predictions and test the sensitivity of total metallicity
and stellar M/L to [$\alpha$/Fe] variations.

Estimates of element abundance ratios for large samples of both early-type and star-forming galaxies
(exploiting the better coverage of the models to high temperatures) will allow us to address several key
questions on SF timescale in galaxies.



\begin{acknowledgments}
\small{
I would like to thank in particular Anna Pasquali and Jakob Walcher for their substantial contribution to
the work presented here. I am grateful to many colleagues for their support in different stages of this
research: St{\'e}phane Charlot, Jarle Brinchmann, Paula Coelho, Gustavo Bruzual, Eric Bell and Simon White.}

\end{acknowledgments}

\end{document}